\documentclass[aps,prd,twocolumn,preprintnumbers,groupedaddress,nofootinbib]{revtex4}
\usepackage{amsmath}
\usepackage{amsfonts}
\usepackage{amssymb}
\usepackage{mathrsfs}
\usepackage{graphicx}
\usepackage{multirow}
\usepackage{color}
\usepackage{hyperref}
          \newcommand{\etal}{{\it et al. }}
          
           \newcommand{\GeV}{{\mathrm {GeV}}}

   \newcommand{\MeV}{{\mathrm {MeV}}}
     \newcommand{\eV}{{\mathrm {eV}}}
\newcommand{\half}{\frac{1}{2}}    \newcommand{\fb}{{\mathrm {fb}}}
\newcommand{\dd}{\mathrm{d}}   \newcommand{\TeV}{{\mathrm {TeV}}}
\newcommand{\mb}[1]{\boldsymbol{#1}}
 \newcommand{\rep}[1]{\mathbf{#1}}
\newcommand{\conjrep}[1]{\overline{\mathbf{#1}}}

\begin{document}
\title{A sharp 141 GeV  Higgs prediction from environmental selection}
\author{James Unwin}
\email{unwin@maths.ox.ac.uk}
\affiliation{Mathematical Institute, University of Oxford,
24-29 St Giles', Oxford, OX1 3LB, UK\\
Rudolf Peierls Centre for Theoretical Physics, University of Oxford,
1 Keble Road, Oxford, OX1 3NP, UK}
\preprint{OUTP-11-54P}

\begin{abstract}
We  construct an environmentally selected supersymmetric standard model with a single Higgs doublet, in  analogy with the work of Hall and Nomura. The low energy spectrum presents only the standard model states with a single Higgs and TeV scale gauginos. The model features a precise Higgs mass prediction $m_H=141\pm2$ GeV and the neutral wino provides a viable dark matter candidate.
\end{abstract}

\maketitle

\section{Introduction}
Recent results from ATLAS \cite{ATLAS} and CMS \cite{CMS} based on luminosities of 1 - 2.3 $\fb^{-1}$ suggest a possible Higgs resonance around 143 GeV.  Notably, Hall \& Nomora, \etal \cite{Elor:2009jp} predicted a Higgs boson mass of $m_H=141\pm2$ GeV in a class of supersymmetric standard models with environmentally selection (E-SSM),  for large $\tan\beta$. This precise prediction of the mass of the Higgs boson was obtained by matching the Higgs quartic coupling $\lambda_H$ with the  supersymmetric boundary condition
\begin{equation}
\lambda_H=\half(g^2+g^{\prime}{}^2)\cos^22\beta
\end{equation}
and using renormalisation group methods to scale the couplings from the SUSY breaking scale to the weak scale. 
The E-SSM is built on the premise that the hierarchy problem is resolved through environmental selection on the scale of electroweak symmetry breaking \cite{hier}. Further, it is argued that whilst supersymmetry (SUSY) is no longer needed to solve the hierarchy problem, it should be present at some scale since it is crucial in any physical realisation of  string theory. Indeed, anthropic arguments occur quite naturally in the context of the string theory landscape and have previously been used to suggest resolutions to a variety of problems, most prominently the cosmological constant \cite{Weinberg:1987dv}. 

We shall most closely emulate the formulation of the E-SSM in which the low energy spectrum contains only the known standard model states, a single Higgs boson and TeV scale gauginos\footnote{This corresponds to scenario IV in  \cite{Elor:2009jp}.}. The other states in the theory acquire masses near the SUSY breaking scale $\tilde{m}$. It can be arranged that the neutral wino is the lightest supersymmetric partner (LSP) and provides a suitable weakly interacting massive particle (WIMP) dark matter (DM) candidate. Alternative formulations with different low energy spectra lead to deviations in some predictions, however, if the DM candidate is replaced by a light singlino, or axion, the Higgs mass for large $\tan\beta$ remains unaltered.

We propose that the simplest manner in which to realise a single scalar Higgs in the IR theory and obtain the precise prediction $m_H=141\pm2$ GeV is through the recently proposed Supersymmetric One Higgs Doublet Model (SOHDM) \cite{Davies:2011mp}. To accommodate high scale SUSY and gauge couplings unification we shall modify some details of the original SOHDM. The resultant model, which we shall refer to as E-SOHDM, has a similar low energy spectrum to the E-SSM by construction and we inherit the Higgs mass prediction of \cite{Elor:2009jp} for the  limit $\tan\beta\rightarrow\infty$. Furthermore, the model naturally provides suitable values for the DM relic density and the neutrino masses.

Split  SUSY models \cite{split}  are closely related to E-SSM. The frameworks differ mainly in their low energy spectra. In particular, motivated by precision unification, models of split SUSY generically have weak scale Higgsinos, due to a small $\mu$ term. E-SODHM is comparable to split SUSY in the limit $\tan\beta\rightarrow\infty$, with $\mu$ at the SUSY breaking scale $\tilde{m}$. Some aspects of split SUSY models with large $\mu$ were previously studied in \cite{Cheung:2005ba}.

This paper is structured as follows, we shall first construct a supersymmetric one Higgs doublet model with high scale SUSY breaking, neutrino masses and wino DM. Following this we shall analyse the low energy spectrum of the model and compare with \cite{Elor:2009jp} to obtain predictions for the Higgs mass. Finally, we shall discuss some alternative formulations of E-SOHDM, in particular the introduction of Dirac gaugino masses, and comment on methods for distinguishing different models of high scale SUSY.

\section{E-SOHDM}
\label{sec2}

Following E-SSM, we assume that the hierarchy problem is resolved through fine-tuning due to environmental selection and further that anthropic requirements also determine the DM relic density. A neutral wino LSP provides a favourable WIMP candidate for DM. The scenario which is most naturally realised in this model is a neutral wino LSP which is nearly degenerate with the charged winos. The consequence of this near degeneracy is that the wino annihilation cross section is Sommerfeld enhanced and this causes a reduction in the wino thermal relic abundance  \cite{Hisano:2006nn}. In order to reproduce the observed DM relic density, at $2 \sigma$, the wino mass must lie in the range:
\begin{equation}
2.5\,\TeV\lesssim M_2 \lesssim 3.0\, \TeV.
\label{DM}
\end{equation}
%
%
%
%
\begin{table}[t]
\begin{center}
\def\str{\vrule height12pt width0pt depth7pt}
\begin{tabular}{| c | l | c | c | }
    \hline\str
    ~Field ~&~ Gauge rep. ~&~ $U(1)_R$ ~&~ $(-1)^{3(B-L)}$~ \\
    \hline\str
    ~~$\mb{Q}$ & ~~$\,(\rep{3},\rep{2})_{1/6}$ & $1$ & $-$\\
    \hline\str
    ~~$\mb{U^c}$ & ~~$\,(\conjrep{3},\rep{1})_{-2/3}$ & $1$ & $-$ \\
    \hline\str
    ~~$\mb{D^c}$ & ~~$\,(\conjrep{3},\rep{1})_{1/3}$ & $1$ & $-$\\
    \hline\str
    ~~$\mb{L}$ & ~~$\,(\rep{1},\rep{2})_{-1/2}$ & $1$ & $-$ \\
    \hline\str
    ~~$\mb{E^c}$ & ~~$\,(\rep{1},\rep{1})_1$ & $1$ & $-$\\
    \hline\str
    ~~$\mb{H}$ & ~~$\,(\rep{1},\rep{2})_{1/2}$ & $0$ & $+$ \\
    \hline\str
    ~~$\mb{\eta}$ & ~~$\,(\rep{1},\rep{2})_{-1/2}$ & $2$ & $+$\\
      \hline\str
    ~~$\mb{X}$ & ~~$\,(\rep{1},\rep{1})_0$ & $2$ & $+$\\
    \hline
\end{tabular}
\caption{Spectrum of chiral superfields.}
\label{Tab1}
\end{center}
\end{table}
%
%
%
Note that without Sommerfeld enhancement the required wino mass range is $1.9\,\TeV\lesssim M_2 \lesssim 2.3\, \TeV$. Whilst it is not possible to observe 2.5 TeV winos at current direct detection experiments, current indirect detection projects and next generation direct detection experiments may be able to test this prediction. The experimental signals are discussed in \cite{Elor:2009jp}.

Having assumed that the weak scale and the DM density are determined by environmental selection, we draw our motivation for model building from what are widely considered the likely properties of high energy physics, neutrino masses, gauge coupling unification and (high scale) supersymmetry.  Furthermore, since we wish to have one Higgs boson in the low energy spectrum we shall insist that only  a single Higgs field is present in the model.

Conventionally, two Higgs doublets are required in (minimal) supersymmetric theories in order to give masses to the quarks and leptons, and to ensure anomaly cancellation. In contrast, a single Higgs doublet, the scalar component of  $\mb{H_u}$, can provide masses to all of the standard model fermions \cite{Davies:2011mp}, \cite{OHD}. The chiral superfield $\mb{H_d}$ is included to cancel anomalies, although it does not obtain a vacuum expectation value (VEV) and symmetries forbid Yukawa couplings involving $\mb{H_d}$. The field $\mb{H_d}$ can no longer be considered a Higgs and consequently is relabelled $\mb{\eta}$. The field $\mb{H_u}$, being the only true Higgs field, is labelled $\mb{H}$. The field content and charges of the chiral superfields are summarised in Table \ref{Tab1}. The chiral superfield $\mb{X}$ is a spurion field which parameterises the SUSY breaking. The model has an anomaly free $U(1)$ $R$-symmetry and matter parity in order to stabilise the LSP. Note that the symmetries forbid Majorana gaugino mass terms and trilinear $\mathcal{A}$-terms, but allow the $\mu$ term. The Higgs sector Lagrangian is given by
\begin{equation}
\begin{aligned}
\mathcal{L}_{\mathrm{H}}=\int\mathrm{d}^2\theta 
\mu\mb{H\eta}
+\int\mathrm{d}^4\theta & \frac{\mb{X^{\dagger}}}{M_*}
 \lambda_\mu\mb{H\eta}.
\label{higgs}
\end{aligned}
\end{equation}
However, since we do not require the Higgsinos to lie near the weak scale, there is no $\mu$-problem. Note that the scale of $\mu$ is the main difference between these high scale SUSY models and split SUSY. The size of $\mu$  leads to deviations in the Higgs mass predictions between the two frameworks. Moreover, the lightest neutralino in models of split SUSY is an unknown mixture of the neutral Higgsinos and gauginos \cite{Wang:2005kf}. Since $\mu\sim\tilde{m}$ in E-SSM and E-SOHDM, the lightest neutralino is almost completely wino and (assuming that this state is responsible for the DM density) this results in a much sharper prediction of the DM mass.

\begin{figure}[t]
\includegraphics[height=55mm,width=75mm]{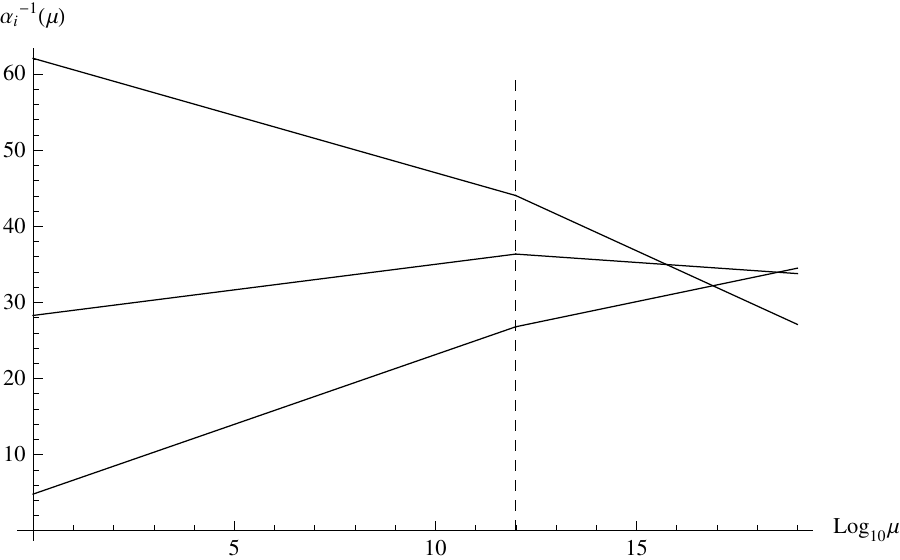}
\caption{
Unification of gauge couplings. The dashed line indicates the scale $\tilde{m}$ at which scale all states contribute to the running.}
\label{Fig1}
\end{figure}

The  $U(1)$ $R$-symmetry forbids Majorana gaugino masses. However, as in split SUSY, suitable gaugino masses can be generated via the model-independent contribution from anomaly mediation \cite{Giudice:1998xp}, \cite{Randall:1998uk} 
\begin{equation}
M_i=\frac{g^2_i}{16\pi^2}b_0^im_{3/2},
\end{equation}
where $b_0^i$ are the $\beta$ function coefficients and $m_{3/2}$ is the gravitino mass. Note that as the gauginos are the only new states introduced below $\tilde{m}$, the $\beta$ function coefficients are similar to the standard model:  
\begin{equation}
b_0^i=\left(-\frac{41}{10},\,\frac{11}{6},\,5\right).
\label{b0}
\end{equation}
Consequently, below $\tilde{m}$ the running of the coupling constants is comparable to the standard model. Gauge unification occurs above the scale $\tilde{m}$ around $M_U\sim10^{17\pm1}$ GeV, with similar precision to the standard model gauge unification, see Figure \ref{Fig1}. From Equation (\ref{b0}) we can obtain the gaugino masses
\begin{equation}
\begin{aligned}
|M_1|&
=5.5\times 10^{-3} m_{3/2},\\
|M_2|&
=5\times 10^{-3} m_{3/2},\\
|M_3|&
=4.3\times 10^{-2} m_{3/2}.
\label{w}
\end{aligned}
\end{equation}
For high scale SUSY breaking $M_{1,2}\gg M_W$ and the absolute values of the $M_i$ correspond very well to the masses of the physical gauginos. The neutral wino is the LSP and is nearly degenerate with the charged winos, a mass splitting of $165$ MeV is generated by electroweak  corrections. Therefore, the wino annihilation cross section is Sommerfeld enhanced, as anticipated. For a gravitino mass $m_{3/2}\sim$ 500 - 600 TeV, the magnitude of $M_2$ from anomaly mediation coincides with the observed DM abundance (\ref{DM}):
\begin{equation}
\begin{aligned}
|M_1|&
\approx 3\,\TeV,\\
|M_2|&
\approx 2.75\,\TeV,\\
|M_3|&
\approx 24\,\TeV.
\label{w}
\end{aligned}
\end{equation}
The gravitino mass is given by
\begin{equation}
 m_{3/2}=\frac{F_X}{\sqrt{3}M_{\mathrm{Pl}}}.
\label{3/2}
 \end{equation}
 where $M_{\mathrm{Pl}}=2.4\times10^{18}$ GeV is the reduced Planck mass. Thus to obtain a suitable gravitino mass we require that the SUSY breaking scale is
\begin{equation}
F_X\approx 4\times10^{24}\GeV.
\label{FX}
 \end{equation}

The standard model fermion masses arise from the following Yukawa terms
\begin{equation}
\begin{aligned}
\mathcal{L}_{\mathrm{Y}}=&\int\mathrm{d}^2\theta \lambda_U \mb{HQU^c}\\
&+\int\mathrm{d}^4\theta
\frac{\mb{X^{\dagger}}\mb{H^{\dagger}}}{M_*^2}\left( \lambda_D \mb{QD^c}+ \lambda_E \mb{QE^c}\right).
\end{aligned}
\end{equation}
All of the quarks and leptons acquire their masses from the VEV of $H$, the scalar component of $\mb{H}$,
\begin{equation}
\langle H\rangle=v/\sqrt{2}\approx 174\,\GeV.
\label{con2}
\end{equation}
To obtain the observed masses, the following tree-level relationship must be satisfied:
\begin{equation}
\frac{\lambda_b F_X}{M_*^2}\times174\GeV \approx 5\,\GeV.
\label{con}
\end{equation}
Hence, by Equation (\ref{FX}), in order to obtain the correct DM relic abundance  we require 
\begin{equation}
M_*\lesssim 5\times 10^{13}\GeV.
\label{con}
\end{equation}

The scale $M_*$ naturally provides suitable neutrino masses through the dimension 5 Weinberg operator
\begin{equation}
\begin{aligned}
\mathcal{L}_{\nu}=&\int\mathrm{d}^4\theta 
\frac{\mb{X^{\dagger}}}{M_*^3} \mb{H^2L^2}.
\end{aligned}
\end{equation}
This term leads to neutrino masses of the order
\begin{equation}
M_{\nu_L}\sim \frac{F_X v^2}{2M_*^3}.
\end{equation}
To accommodate the observed neutrino scale we require that
$ 5.8\times10^{-2}\, \eV \lesssim \sum_{\nu}m_{\nu}\lesssim 0.28\, \eV$ \cite{neutrinos}.
By comparison with Equation (\ref{con}), and assuming natural couplings $0.1<\lambda_b<1$,  we obtain
\begin{equation}
5\times 10^{12}\GeV< M_*<  5\times 10^{13}\GeV,
\end{equation}
and it is clear that phenomenological acceptable neutrino masses can be generated
\begin{equation}
5\times 10^{-4}\eV\lesssim M_{\nu_L}\lesssim 0.5 \,\eV.
\end{equation}
From an anthropic perspective, neutrino masses much higher then this greatly suppress structure formation due to free streaming. For a combined neutrino mass larger than $\sum_{\nu}m_{\nu}\gtrsim 10\, \eV$ it is found that galaxy formation is strongly suppressed (by a factor greater than $10^{-4}$) \cite{Tegmark:2003ug}. This presents a catastrophic boundary in the landscape and an anthropic constraint on the relative magnitudes of the SUSY breaking and cutoff scales. 
\begin{equation}
M_* \gtrsim 3 \times 10^{12} \GeV.
\end{equation}
It is likely that the SUSY breaking scale $\tilde{m}$ and the UV cutoff $M_*$ are comparable, $\sim 10^{12\pm1}$GeV, and related. In the context of string theory the compactification scale can provide a suitable UV cutoff
\begin{equation}
M_*^{D-2}\sim\frac{M_{\mathrm{Pl}}^2}{\mathcal{V}}.
\end{equation}
With a large compactification volume $\mathcal{V}$ we can obtain a suitable  $M_*\ll M_{\mathrm{Pl}}$.

The models $U(1)$ $R$-symmetry should viewed as an emergent symmetry of the low energy theory. To resolve the cosmological constant problem the $R$-symmetry must be broken at high scales by supergravity effects and this results in an $R$-axion. Since the SUSY breaking scale is high the $R$-axion is heavy
\begin{equation}
m_a^2\sim\frac{|F_X|^{3/2}}{M_{\mathrm{Pl}}}\sim 10^{18}\GeV^2.
\end{equation}
Consequently, the  $R$-axion, and likewise the gravitino $m_{3/2}\sim500$ TeV, are heavy enough to evade all cosmological constraints and searches \cite{Raxion}. Whilst the scale of the SUSY breaking is not sufficiently high in order to avoid all cosmological problems due to moduli (specifically, the modulus field associated to the overall volume has a mass $\sim1$ GeV), discussions on circumventing these difficulties can be found in  \cite{Conlon:2007gk}.

\section{Higgs Mass prediction}

\begin{figure}[t]
\includegraphics[height=55mm,width=75mm]{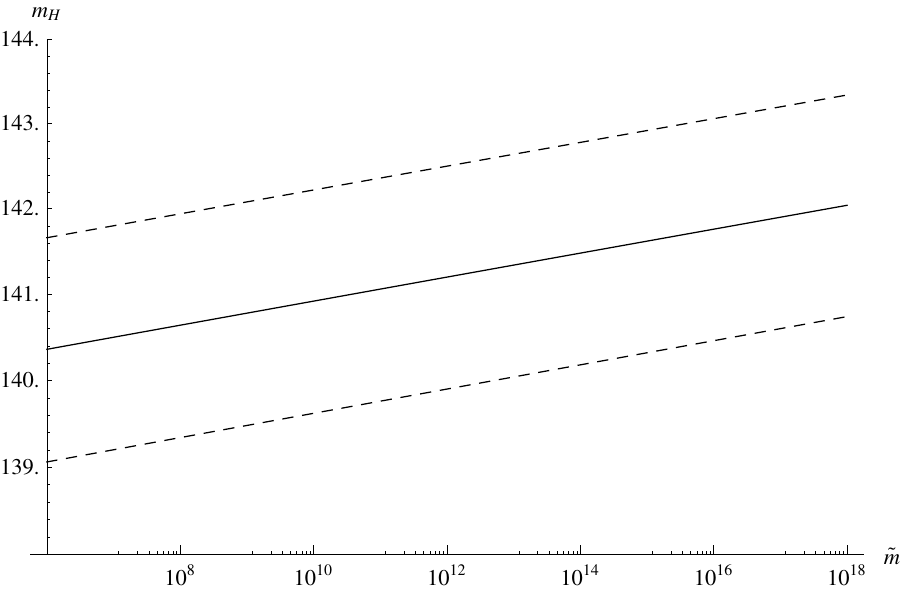}
\caption{
Higgs mass $m_H$ plotted against $\tilde{m}$ (GeV). The dashed upper (lower) line displays the effect of increasing (decreasing) the top mass by $0.9$GeV. We take $\alpha_S(M_Z)=0.1184$.}
\label{Fig2}
\end{figure}

By construction, the spectrum below $\tilde{m}$ is relatively unchanged from the E-SSM. Since the gluinos and bino do not couple directly to the standard model Higgs boson the mass prediction is unchanged from the calculation with only light winos presented in \cite{Elor:2009jp}.  The low energy Higgs potential may be written in the form
\begin{equation}
V_H=
-\frac{m_H^2}{2}|H|^2+\frac{\lambda_H}{4}|H|^4,
\end{equation}
where $m_H$ the physical Higgs boson mass and is given by
\begin{equation}
m_H^2=\frac{\lambda_Hv^2}{2}.
\end{equation}
The VEV $\langle H \rangle$ may be expressed
\begin{equation}
 \frac{v}{\sqrt{2}}=2\sqrt{\frac{\tilde{m}_H^2-|\mu|^2}{g^2+g^{\prime}{}^2}}\approx174 \,\GeV,
\end{equation}
where $\tilde{m}_H$ is the Higgs soft mass. The value of the Higgs mass may be calculated by noting that  the quartic Higgs coupling is fixed by the supersymmetric boundary condition at the scale $\tilde{m}$
\begin{equation}
\lambda_H(\tilde{m})=\frac{g^2(\tilde{m})+g^{\prime}{}^2(\tilde{m})}{2}(1+\delta(\tilde{m})),
\label{ssbc}
\end{equation}
where the quantity $\delta$ accounts for threshold corrections at the scale $\tilde{m}$. Convergence of the IR flow makes the Higgs mass prediction relatively insensitive to $\delta$. Hall \& Nomura used renormalisation group scaling to run all of the couplings from $\tilde{m}$ to the weak scale in order to determine the physical Higgs mass. In their analysis they included one loop weak scale threshold corrections (including the winos), and two and three loop QCD effects. The main uncertainties on the mass prediction come from errors on the top mass $m_t$ and QCD coupling $\alpha_S(M_Z)$.  The current experimental values for these parameters are \cite{Tevatron},  \cite{Bethke}:
\begin{equation}
\begin{aligned}
m_t &=173.1\pm0.9 \,\GeV,\\
\alpha_S(M_Z) &=0.1184\pm0.0007.
\end{aligned}
\end{equation}
The Higgs mass $m_H$ prediction is similar to that of E-SSM (with wino DM) in the limit $\tan\beta\rightarrow\infty$. Here we recapitulate the relevant result of \cite{Elor:2009jp} with updated errors:
\begin{equation}
\begin{aligned}
m_H=&141.2\,\GeV \\
&+1.3\, \GeV \left(\frac{m_t-173.1 \,\GeV}{0.9\, \GeV}\right)\\
&-0.35\, \GeV \left(\frac{\alpha_s(M_Z)-0.1176}{0.0007}\right)\\
&+0.14\, \GeV \log_{10}\left(\frac{\tilde{m}}{10^{12} \,\GeV}\right).
\label{eqn}
\end{aligned}
\end{equation}
Removing the uncertainty introduced by $\tan\beta$, inherent to E-SSM, leads to this sharp Higgs mass prediction in E-SOHDM. Remarkably, order of magnitude changes to $\tilde{m}$ result in only small ($\sim100\, \MeV$) deviations in the Higgs mass prediction, as can be seen in Figure \ref{Fig2}. As the errors on the top mass and QCD coupling shrink the leading uncertainty will come from the last term in Equation (\ref{eqn}). Thus information on the scale $\tilde{m}$ can be obtained from precision measurements of the Higgs mass.

\section{Dirac Gauginos}
\label{sec4}

\begin{table}[t]
\begin{center}
\def\str{\vrule height12pt width0pt depth7pt}
\begin{tabular}{| c | l | c | c | }
    \hline\str
    ~Field ~&~ Gauge rep. ~&~ $U(1)_R$ ~&~ $(-1)^{3(B-L)}$~ \\
    \hline\str
    ~~$\mb{T}$ & ~~$\,(\rep{1},\rep{3})_0$ & $0$ & $+$\\
      \hline\str
    ~~$\mb{O}$ & ~~$\,(\rep{8},\rep{1})_0$ & $0$ & $+$\\
      \hline\str
    ~~$\mb{S}$ & ~~$\,(\rep{1},\rep{1})_0$ & $0$ & $+$\\
    \hline\str
    ~~$\mb{W^{\prime}}$ & ~~$\,(\rep{1},\rep{1})_0$ & $0$ & $+$ \\
    \hline\str
    ~~$\mb{B}$ & ~~$\,(\rep{3},\rep{2})_{-5/6}$ & $0$ & $+$ \\
    \hline\str
    ~~$\overline{\mb{B}}$ & ~~$\,(\rep{3},\overline{\rep{2}})_{ 5/6}$ & $0$ & $+$ \\
    \hline
\end{tabular}
\caption{Additional chiral superfields for models with Dirac gaugino masses.}
\label{Tab2}
\end{center}
\end{table}

Motivated by minimality, in the model of Section \ref{sec2} we did not include extended superpartner (ESP) fields, to provide Dirac masses to the gauginos,  as in the original SODHM. Supplementing the spectrum in Table \ref{Tab1} with chiral superfields $\mb{O}$, $\mb{T}$ and  $\mb{S}$, detailed in Table \ref{Tab2}, one can construct Dirac mass terms for the gauginos  \cite{Dirac}-\cite{Abel:2011dc}. Generically, this leads to the mass hierarchy \cite{Benakli:2010gi}:
\begin{equation}
\begin{aligned}
m_\lambda &\sim \frac{g\lambda}{16\pi^2}\frac{D^{\prime}}{M_*},\\
m_{\tilde{f}} &\sim \frac{g^2}{16\pi^2}\frac{F_X}{M_*},\\
m_{\phi} &\sim \frac{\lambda}{4\pi}\frac{D^{\prime}}{M_*}, \quad \frac{\lambda}{4\pi}\frac{F_X}{M_*},
\end{aligned}
\end{equation}
where $m_{\lambda}$ and $m_{\phi}$ are the gaugino and scalar ESP masses, respectively.
Let us suppose that the low energy spectrum of the model contains the standard model, the gauginos and the ESP states. To obtain a suitable splitting in the spectrum we shall assume that   
$D^{\prime}, F_X\sim\tilde{m}M_*$
 and that $\lambda_2$ is tuned small through environmental selection on the mass of the wino DM. Models with comparable  $D$- and $F$- term breaking have been studied in \cite{SB}.  With natural couplings $\lambda_1,\, \lambda_3\sim 1$ the Dirac binos and gluinos, and the associated scalar ESPs have masses $\sim\tilde{m}$. The only additions to the SM  states at low energy are the Dirac winos and the associated scalar ESP and this will not cause a large deviation in the Higgs prediction as the couplings to the Higgs field are at higher order
\begin{equation}
\int\dd^4\theta\frac{\mb{X^{\dagger}}}{M_*}\Big(\lambda_S \mb{SH\eta} + \lambda_T \mb{HT\eta}\Big).
\end{equation}

The ESP fields can be embedded into an adjoint representation of $SU(5)$. The adjoint of $SU(5)$ may be decomposed over the standard model gauge groups as follows:
\begin{equation*}
\mb{24}=(\rep{8},\rep{1})_0+(\rep{1},\rep{3})_0+(\rep{1},\rep{1})_0+(\rep{3},\rep{2})_{-5/6}+(\rep{3},\overline{\rep{2}})_{ 5/6}
\end{equation*}
In order to complete the adjoint representation we introduce a pair of vector-like `bachelor' superfields $\mb{B}$ and  $\overline{\mb{B}}$ with masses at the scale $\tilde{m}$. 
These new fields alter the $\beta$ function coefficients, however there is still approximate gauge coupling unification. The unification scale depends strongly on the magnitude of $\tilde{m}$. 
There is a danger that the hypercharge ESP singlet field $\mb{S}$ may acquire a large tadpole term, however this can be avoided if the couplings to the messengers respect $SU(5)$ \cite{Abel:2011dc} or are otherwise suitably arranged \cite{Benakli:2010gi}. 

We now construct the following Dirac mass terms: 
\begin{equation}
\begin{aligned}
\Delta\mathcal{L}=\int\dd^2\theta\frac{\mb{W^{\prime}_\alpha}}{M_*}\Big(&\lambda_3 \mathrm{Tr}(\mb{OW_3^{\alpha}})\\
&+\lambda_2 \mathrm{Tr}(\mb{TW_2^{\alpha}})+\lambda_1\mb{SW_1^{\alpha}}\Big)
\end{aligned}
\end{equation}
where $\mb{W_i}$ are the gauge superfields of the standard model gauge groups and $\mb{W^{\prime}}$ is a spurious $U(1)^{\prime}$ gauge superfield.  This leads to gaugino mass terms of the form
\begin{equation}
M_3 \mathrm{Tr}(\tilde{O}\tilde{G}) + M_2 \mathrm{Tr}(\tilde{T}\tilde{W}) + M_1 \mathrm{Tr}(\tilde{S}\tilde{B})
\end{equation}
where
\begin{equation}
M_i=\frac{\lambda_i D^{\prime}}{M_*}.
\label{Mi}
\end{equation}
The mass hierarchy can be suitably arranged such that the neutral wino is the LSP, which is nearly degenerate with the charged winos. From inspection of the wino mass $M_2$, given in Equation (\ref{Mi}), we find that in order to reproduce the observed dark matter density (cf. Equation (\ref{DM})) we require that 
\begin{equation}
 \tilde{m}\sim \frac{3\,\TeV}{\lambda_2}. 
\end{equation}
Comparing this requirement with Equation (\ref{con2}), which ensures suitable standard model Yukawa couplings, leads to the following condition:
\begin{equation}
M_*\gtrsim \frac{100 \,\TeV}{\lambda_2 }.
\end{equation}

To generate the splitting in the spectrum we have previously assumed that $\lambda_2\ll 1$ and thus $M_*$ may be large. This allows for appropriate couplings neutrino masses to be introduced through the Weinberg operator as in Section \ref{sec2}. Employing this mechanism fixes $\tilde{m}$; for $\lambda_2\sim10^{-8}$ and $M_*\sim 10^{13}$ GeV to obtain TeV scale winos the SUSY breaking scale must be $\tilde{m}\sim10^{11}$ GeV. As noted previously, the scale of $\tilde{m}$ could be determined by precision Higgs mass measurements and therefore in principle the origin of the neutrino masses could be tested.

\section{Phenomenology}

We shall make some brief comments on how models with high scale SUSY could be probed and in particular methods by which E-SODHM can be distinguished from alternative models.
All models with exceptionally heavy sfermions avoid flavour problems and limits from proton stability are relaxed. Whilst the dimension 5 operators $\mb{Q}\mb{Q}\mb{Q}\mb{L}$ and $\mb{U^c}\mb{U^c}\mb{D^c}\mb{E^c}$ can still lead to dangerous four-fermion vertices, in E-SODHM these operators are forbidden by the $U(1)$ $R$-symmetry. Indeed, the $U(1)$ $R$-symmetry is one of the most distinguishing features of E-SOHDM and it has distinctive collider signatures which could be used to differentiate it from other theories of high scale SUSY in the large $\tan\beta$ limit. 

 If the gauginos have Dirac masses and the Majorana masses are forbidden by the $U(1)$ $R$-symmetry, this alters the available production and decay channels of gauginos and sfermions. Whilst most of these states lie beyond the reach of the current technology, indirect searches may be possible. In particular, an analysis of the ratio of like to unlike sign di-lepton events with large missing energy could potentially determine the nature of the gauginos \cite{Choi:2008pi}. As the sfermions are generally very heavy this analysis would require a large amount of data and careful study.
Moreover, if the ESP fields, considered in Section \ref{sec4}, are present then this can lead to distinct signals which could be used to distinguish this model from alternative proposals. The phenomenology of the ESP sector has been studied in \cite{ESP}.

Whilst it is not inconceivable that the effects of the TeV scale gauginos could be detected in a next-generation collider, perhaps a more immediately accessible window on models with environmental selection is provided by observations of the early universe. In particular, deviations during Big Bang nucleosynthesis (BBN). 
In E-SOHDM the bino has no direct decay route and must first mix with the Higgsino and the gluinos can only decay via heavy squarks. Consequently, both binos and gluinos have long lifetimes and their late decays could result in observable signals during BBN. The effects of decaying gluinos during BBN have been consider in the context of split SUSY \cite{split} and general constraints on energy injection during BBN are studied in \cite{BBN}. These cosmological constraints can be ameliorated if one assumes that the reheating temperature is less than the gluino/bino mass, such that these states can not be produced after reheating.

\section{Concluding remarks}
The two  main predictions of E-SOHDM are the Higgs mass $m_H\approx 141$ GeV and the DM mass $m_{\chi}\approx 2.75$ TeV. The Higgs mass prediction is much sharper, in comparison to the general prediction of the E-SSM with wino DM:
\begin{equation}
127\, \GeV\lesssim m_H\lesssim 142 \, \GeV.
\end{equation}
Here we have presented what we consider to be the most aesthetic formulation of an environmentally selected SOHDM. However, there are several interesting alternatives which could be constructed. 
GUT groups other than $SU(5)$ could be considered  and this would have an effect on the Higgs mass prediction. For instance, the simplest implementation of $SO(10)$ unification causes the mass prediction to shift by $\delta M_H\approx2.4$ GeV \cite{Elor:2009jp}.

In the construction of E-SOHDM we have imposed a discrete symmetry in order to stabilise the LSP, such that it can provide a viable DM candidate. Without this symmetry the most natural alternative is to assume that the axion is responsible for the DM relic abundance. Anthropic arguments for axion DM have been discussed at length in literature \cite{axion}.

In specific models the SUSY breaking scale may be fixed by phenomenological requirements (such as obtaining suitable neutrino masses via the Weingberg operator, as in Section \ref{sec4}) or aesthetics (assuming $M_*\sim \tilde{m}$, as in section \ref{sec2}).  However, in a variety of models $\tilde{m}$ could in principle lie anywhere between the TeV scale and the Planck scale.  If $\tilde{m}$ is relatively low then this can lead to cosmological signals from the gravitino, $R$-axion and moduli \cite{Raxion}, \cite{Conlon:2007gk}. Notably, as can be seen from Figure \ref{Fig2}, the Higgs mass prediction is relatively insensitive to changes in $\tilde{m}$.

There are a number of alternative high scale SUSY models in the literature which have comparable Higgs mass predictions  \cite{split}, \cite{Other}. Most prominently, models of split SUSY have both similar low energy spectra and  Higgs mass predictions  \cite{split}:
\begin{equation}
110\, \GeV \lesssim m_H \lesssim 156 \,\GeV.
\end{equation}
 If the symmetry structure of E-SOHDM was altered to ensure a weak scale $\mu$-term, this would lead to models similar to split SUSY. The inclusion of weak scale Higgsinos alters both the dark matter  \cite{Wang:2005kf} and Higgs mass predictions  \cite{split}. From existing analysis, the Higgs boson mass in split SUSY with one Higgs doublet would be $\sim140$ GeV, with the main uncertainty coming from the magnitude of $\tilde{m}$. 

Whilst supersymmetry or new strong dynamics could ultimately resolve the hierarchy problem, if only the standard model Higgs is found after the full LHC run with $100\,\fb^{-1}$ then more radical ideas must be seriously contemplated.  If the LHC discovers a single Higgs at $m_H\approx141$ GeV and no signals of physics beyond the standard model, then it becomes highly plausible that fine-tuning is inherent to the physical universe. The existence of such fine-tuning would lend exceptional credence to the concept of the  multiverse.

\section*{Acknowledgements} 
I am grateful to John March-Russell for useful discussions and Matthew McCullough for detailed comments on the manuscript. This work was partially funded by an EPSRC doctoral studentship and awards from St. John's College \& Pembroke College, Oxford.

\end{document}